# Religion and Artificial Intelligence as Distributed Meaning Systems: A Naturalistic Conceptual Model

Dhushy Thillaivasan
School of Business, Melbourne Institute of Technology, Melbourne, Australia
Email: DThillaivasan@mit.edu.au
*Corresponding author

Samar Shailendra
School of IT & Engineering, Melbourne Institute of Technology, Melbourne, Australia
Email: sshailendra@mit.edu.au

Sohag Sarkar
School of Business, Melbourne Institute of Technology, Melbourne, Australia
Email: ssarkar@academic.mit.edu.au

## Abstract

This paper develops a naturalistic account of religion and artificial intelligence as structurally similar distributed meaning systems. I argue that both emerge from the same underlying cognitive architecture: socially extended processes that offload interpretation, norm-guidance, and world-model construction into external symbolic environments. Drawing on work in distributed cognition, cultural evolution, and philosophy of mind, the paper proposes a conceptual model showing how meaning is generated, stabilised, and transmitted through recursive interactions between agents and their informational ecologies.

Religion is analysed not as a set of beliefs but as a cognitive-ecological system that scaffolds coordination, normativity, and shared interpretation. Contemporary AI systems are shown to instantiate analogous functions, operating as high-bandwidth, algorithmically mediated environments that shape reasoning, attention, and social meaning-making.

The model explains how both systems create epistemic compression, reduce cognitive load, and generate shared frameworks that guide behaviour. It also clarifies the conditions under which AI systems can become culturally entrenched meaning authorities.

The contribution is conceptual: a unified framework for understanding religion and AI as parallel forms of distributed cognitive machinery. This reframing opens new pathways for analysing artificial agents not as isolated tools but as components in evolving socio-cognitive ecologies.

## 1. Introduction

Human societies have long relied on systems that scaffold interpretation, coordinate behaviour, and stabilise shared understandings of the world. Religion is one of the most enduring of these meaning-making architectures. Artificial intelligence (AI), though emerging from a radically different substrate, is increasingly implicated in similar cognitive-ecological processes: shaping attention, guiding decisions, structuring informational environments, and influencing how individuals and groups construct shared significance. As AI systems become embedded in everyday cognition, they participate in meaning-making dynamics traditionally associated with cultural and institutional formations. Yet despite growing interest in the philosophical and cognitive implications of AI, there remains no naturalistic, conceptually unified account that examines religion and AI together as structurally comparable distributed meaning systems.

Existing scholarship tends to approach the relationship between religion and AI from three directions. The first is theological or normative, asking whether AI challenges or extends religious doctrines. The second is anthropological, examining how people imagine, ritualise, or mythologise AI (Singler, 2020). The third is speculative, projecting future scenarios in which AI becomes an object of worship or a quasi-divine authority. While each of these perspectives offers valuable insights, none provides a naturalistic, cognitive-ecological framework capable of explaining why religion and AI exhibit parallel patterns of emergence, authority, and coordination. What is missing is a conceptual vocabulary that can analyse both domains as distributed, evolving systems that shape human interpretation and behaviour through externalised cognitive scaffolding.

This paper develops such a framework. Drawing on Dennett's account of religion as a natural phenomenon (Dennett, 2006), Harari's theory of intersubjective realities (Harari, 2014), Henrich's work on cultural evolution (Henrich, 2015), Singler's analysis of AI-related myth-making (Singler, 2020), Agüera y Arcas's account of AI as distributed intelligence (Agüera y Arcas, 2022; Aguera y Arcas, 2025), Harari's recent argument that AI is becoming a meaning-shaping system (Harari, 2024a), and Mallaby's account of the emergence of non-biological intelligence (Mallaby, 2026), we argue that both religion and AI can be understood as distributed, evolving systems that generate and regulate shared meaning. This naturalistic perspective does not claim that AI is a religion, nor that religion is reducible to computation. Rather, it identifies structural parallels in how both systems coordinate behaviour, stabilise norms, and produce authoritative interpretations through recursive interactions between agents and their informational environments.

The contribution of this paper is twofold. First, it offers a unified conceptual model that situates religion and AI within a shared analytical vocabulary grounded in five dimensions: evolutionary, cognitive, intersubjective, technological, and anthropological. Second, it demonstrates how this model clarifies emerging dynamics around AI, including the rise of algorithmic authority, the integration of AI into intersubjective meaning structures, and the cognitive-ecological conditions under which AI systems can become culturally entrenched sources of interpretation and coordination.

The paper proceeds as follows. Section 2 outlines naturalistic accounts of religion. Section 3 examines emerging naturalistic accounts of AI as a distributed meaning-making system. Section 4 develops the conceptual framework that aligns these domains. Section 5 presents the unified model. Section 6 analyses structural parallels and divergences. Section 7 discusses the theoretical, methodological, and conceptual implications of the model. Section 8 concludes.

## 2. Naturalistic accounts of religion

Naturalistic approaches treat religion not as a sui generis domain but as a phenomenon explainable through the same mechanisms that account for minds, cultures, and institutions. On this view, religious beliefs, practices, and institutions arise from evolved cognitive dispositions, cultural learning, and social coordination rather than from supernatural revelation (Atran, 2002; Boyer, 2002; Dennett, 2006). This section synthesises three strands of naturalistic work central to our argument: religion as a natural phenomenon, religion as a product of cultural evolution, and religion as an intersubjective reality that structures shared meaning.

### 2.1 Religion as a natural phenomenon

Dennett (2006) argues that religion should be treated as a natural phenomenon open to empirical and theoretical investigation rather than insulated from scrutiny. On this account, religious systems emerge from ordinary cognitive capacities, agency detection, narrative construction, norm tracking that are elaborated through cultural processes. Religious concepts and practices are thus continuous with other human inventions, such as law, music, and money, which also organise behaviour and meaning at scale.

This perspective aligns with the cognitive science of religion, which explains religious beliefs and rituals through domain-general cognitive mechanisms and their cultural elaboration (Atran, 2002; Boyer, 2002; Norenzayan, 2013). Hyperactive agency detection, minimally counterintuitive concepts, and coalitional signalling are not *religious modules* but ordinary features of human cognition that, under particular ecological and social conditions, give rise to religious representations and practices.

Crucially, these accounts already frame religion as a distributed cognitive system. Religious doctrines, rituals, and institutions are not reducible to individual beliefs; they are patterns that persist across generations and are sustained through interactions among people, artefacts, and practices. This aligns with broader accounts of distributed cognition, where functional organisation emerges from the coordinated activity of multiple components rather than from any single locus of understanding (Hutchins, 1995).

### 2.2 Religion, cultural evolution, and collective intelligence

Henrich's work on cultural evolution deepens this naturalistic picture by showing how religious systems emerge through cumulative cultural selection (Henrich, 2015). Human groups rely not on individual insight but on socially transmitted practices, norms, and institutions that have been filtered through historical processes of success and failure. Much of what individuals *know* is embodied in traditions and institutions they follow without fully understanding.

Religion, in this framework, functions as part of a broader ecology of cultural adaptations. Costly rituals, moralising gods, and doctrinal systems stabilise cooperation, enforce norms, and support large-scale social organisation, even when individuals cannot articulate why these practices *work* (Henrich, 2015; Norenzayan, 2013). These insights align with broader cultural evolutionary theory, which emphasises how traditions persist and spread through processes of variation, selection, and retention(Richerson & Boyd, 2005).

This perspective foregrounds two features central to our later comparison with AI. First, religious systems are distributed: no individual contains the full content or logic of the system. Second, they are evolving doctrines, practices, and organisational forms change over time as they are copied, modified, and selected. Religion is therefore a dynamic, historically contingent system of collective intelligence.

### 2.3 Religion as intersubjective reality and meaning system

Harari's notion of *intersubjective realities* provides a complementary lens on religion as a meaning-making system (Harari, 2014). Intersubjective realities are entities, such as nations, corporations, and legal systems, that exist not in individual minds or in the physical world alone, but in the shared beliefs and practices of many people. Religion is a paradigmatic case: gods, sacred texts, and moral orders exert real causal force because they are collectively recognised, narrated, and enacted.

This view aligns with social ontology, which analyses how shared symbolic structures generate institutional facts and normative expectations (Searle, 1995). Religion functions as a technology for generating and stabilising shared meanings: it provides narratives about origins and destinies, frameworks for evaluating actions, and categories for classifying people and events. These narratives and frameworks structure institutions, laws, and everyday practices, shaping how individuals experience themselves and others.

Taken together, these naturalistic accounts converge on a picture of religion as a distributed, evolving system of shared meaning and coordination. It is distributed because its content and causal power are spread across minds, practices, artefacts, and institutions. It is evolving because it changes through cultural transmission and selection. And it is meaning-generating because it provides the categories, narratives, and norms through which people interpret the world and their place within it.

In the remainder of the paper, we extend this naturalistic, system-level understanding to contemporary AI. We do not claim that AI is a religion, nor that religion is reducible to computation. Rather, we argue that both can be analysed as distributed, evolving systems that generate and regulate shared meaning, and that this structural parallel illuminates the emerging role of AI in human social and cultural life.

## 3. Naturalistic Accounts of Artificial Intelligence

Naturalistic approaches to AI treat contemporary systems not as autonomous agents or proto-persons but as products of computational architectures, optimisation processes, and socio-technical environments. On this view, AI systems can be analysed using the same conceptual tools used to study biological cognition, cultural evolution, and distributed information processing. This section synthesises four strands of work central to our argument: AI as emergent non-biological intelligence, AI as distributed intelligence, AI as a meaning-shaping system, and AI as a cultural and mythic object.

### 3.1 AI as emergent non-biological intelligence

Mallaby's analysis of DeepMind provides a naturalistic account of how contemporary AI systems achieve competence without invoking agency, intention, or consciousness (Mallaby, 2026). Modern AI capabilities arise from recursive optimisation, scaling, and feedback-driven improvement rather than from explicit symbolic reasoning or human-like understanding. These systems do not *know* or *intend* in any meaningful sense; instead, they acquire functional competence through iterative training processes that refine performance across vast computational substrates.

This framing parallels naturalistic accounts of biological and cultural evolution, where complex behaviours emerge from selection-like dynamics rather than deliberate design or comprehension (Dennett, 2017; Henrich, 2015). Mallaby's contribution is to show that AI systems exhibit a similar pattern: they become capable not because they possess internal agency but because optimisation processes shape their behaviour over time. This positions AI as a non-biological emergent system whose abilities reflect the structure of training environments and algorithms rather than any intrinsic intentionality.

### 3.2 AI as distributed intelligence

Agüera y Arcas argues that contemporary machine-learning systems exhibit forms of distributed intelligence, in which functional behaviour emerges from the collective dynamics of large-scale networks rather than from any centralised, agent-like locus of control (Agüera y Arcas, 2022; Aguera y Arcas, 2025). On this view, AI systems do not contain stable internal models, semantic representations, or unified *minds*. Instead, they operate as pattern-completion mechanisms whose outputs reflect statistical regularities distributed across training data and computational layers.

This account aligns with broader theories of distributed cognition, which emphasise that cognitive organisation can emerge from interactions among multiple components rather than from a single, fully informed agent (Hutchins, 1995). In both cases, apparent coherence arises from global organisation rather than internal comprehension. Agüera y Arcas therefore positions AI within a broader family of distributed cognitive systems, including cultural traditions, markets, and scientific communities, where no single individual possesses the full content or logic of the system, yet coherent patterns emerge at scale.

For our purposes, this perspective provides the conceptual grounding for treating AI as a system-level phenomenon. Its behaviour is the product of distributed processes that span data, architectures, and computational interactions rather than the expression of an underlying agent. This makes Agüera y Arcas's account essential for understanding how AI can participate in meaning-making without possessing comprehension or intentionality.

### 3.3 AI as a meaning-shaping system

Harari argues that contemporary AI systems are becoming powerful meaning-shaping infrastructures, capable of generating narratives, classifications, and interpretations that influence how people understand themselves and the world (Harari, 2024b). Unlike earlier technologies, which primarily transmitted or stored human-generated meaning, AI systems can now produce text, images, and explanations that are coherent, personalised, and socially resonant. These outputs increasingly mediate how individuals interpret events, evaluate actions, and relate to others.

Crucially, Harari does not attribute beliefs, intentions, or consciousness to AI systems. His argument is that AI participates in meaning-making because humans integrate its outputs into their interpretive practices, not because the systems themselves comprehend what they generate. This aligns with broader analyses of information systems, which emphasise how technological infrastructures shape meaning and social reality (Floridi, 2014).

AI thus functions as a new symbolic layer within human societies: a technology that influences intersubjective realities by generating descriptions, predictions, and stories that people treat as informative or authoritative. For our purposes, Harari's account is essential because it shows how AI can shape shared meaning without possessing agency or understanding, thereby aligning with the naturalistic, system-level approach developed in Section 2.

### 3.4 AI as a cultural and mythic object

Singler's anthropological work demonstrates that AI is also embedded within myth-making practices, where people project agency, intentionality, or even divinity onto computational systems (Singler, 2020). These projections do not arise from the technical properties of AI itself but from cultural narratives that frame AI as a powerful, mysterious, or morally significant entity. Such narratives include fears of superintelligent takeover, hopes for digital immortality, and expectations that AI will deliver objective truth or moral clarity.

These narratives frequently draw on religious tropes, prophecy, transcendence, omniscience, salvation, and apocalypse, thereby positioning AI within familiar symbolic frameworks. They also intersect with broader sociotechnical imaginaries, which shape how societies envision technological futures and the moral significance of emerging systems (Jasanoff & Kim, 2015). These imaginaries influence public understanding of AI, shaping expectations, anxieties, and institutional responses.

For our purposes, Singler's analysis highlights that AI is not only a technical artefact but also a cultural object around which communities construct meaning. These mythic narratives do not describe what AI is; they describe how humans interpret and respond to it. This makes Singler's work essential for understanding how AI becomes entangled in symbolic, ritual, and narrative practices, paralleling the ways religious systems organise meaning and social behaviour.

### 3.5 Synthesis: AI as a distributed, evolving meaning system

Taken together, these naturalistic accounts converge on a picture of AI as a distributed, evolving system whose outputs shape human meaning-making. Mallaby shows that contemporary AI capabilities emerge from iterative optimisation processes rather than from agency or understanding (Mallaby, 2026). Agüera y Arcas demonstrates that AI behaviour arises from distributed network interactions with no centralised locus of control or comprehension (Agüera y Arcas, 2022; Aguera y Arcas, 2025). Harari adds that AI systems increasingly function as meaning-shaping infrastructures, generating narratives and classifications that influence how people interpret themselves and the world (Harari, 2024a). Singler's work shows that AI is also embedded in cultural and mythic narratives that shape public expectations and institutional responses independently of the technology's actual properties (Singler, 2020).

Together, these perspectives support a naturalistic understanding of AI as a system-level phenomenon. Its behaviour is produced by distributed computational processes, refined through evolutionary-like optimisation, and integrated into human interpretive practices in ways that shape shared meaning. AI does not possess beliefs, intentions, or comprehension; rather, it participates in meaning-making because humans treat its outputs as informative, authoritative, or socially significant.

This synthesis parallels the account of religion developed in Section 2. Both religion and AI involve distributed structures, evolutionary dynamics, and meaning-generating processes that operate at the system level rather than the individual level. These structural parallels provide the conceptual foundation for the framework developed in Section 4.

## 4. A Naturalistic Framework for Understanding Religion and AI

The preceding sections established that both religion and AI can be analysed naturalistically as distributed, evolving systems that generate and regulate shared meaning. In this section, we develop a conceptual framework that brings these strands together. Our aim is not to collapse religion into computation or elevate AI into a quasi-religious domain, but to articulate a structural analogy grounded in naturalistic explanations of cognition, culture, and meaning. The framework rests on three core claims: (1) both religion and AI are distributed systems; (2) both evolve through iterative processes of selection and stabilisation; and (3) both participate in the generation and regulation of shared meaning.

### 4.1 Distributed systems

Both religion and AI exhibit distributed organisation, in which no single individual or component contains the full content, logic, or causal power of the system. In religious traditions, meaning is distributed across texts, rituals, institutions, artefacts, and the shared expectations of communities (Dennett, 2017; Henrich, 2015). This aligns with broader theories of distributed cognition, which emphasise that functional organisation emerges from interactions among multiple components rather than from a single locus of understanding (Hutchins, 1995).

Similarly, contemporary AI systems operate through the collective dynamics of networked components, where functional behaviour emerges from statistical patterns distributed across training data and model parameters (Agüera y Arcas, 2022; Aguera y Arcas, 2025). No computational unit “knows” what it is doing; coherence arises from the global organisation of the system rather than from internal comprehension. This distributed structure is central to understanding how both religion and AI generate stable patterns of behaviour and meaning.

### 4.2 Evolving systems

Religion and AI also share an evolutionary character, though the mechanisms differ. Religious systems evolve through cultural transmission, modification, and selection, producing traditions that persist because they solve coordination problems, stabilise cooperation, or resonate with human cognitive dispositions (Henrich, 2015; Norenzayan, 2013). AI systems evolve through optimisation processes, where training regimes iteratively refine performance through feedback, scaling, and environmental interaction (Mallaby, 2026).

In both domains, competence emerges without comprehension. Religious practices often *work* long before participants can articulate why; AI systems achieve high-level performance without possessing understanding or intention. This parallel supports a naturalistic account of how complex, adaptive systems arise from iterative processes rather than from deliberate design.

### 4.3 Meaning-generating systems

The third point of convergence is that both religion and AI function as meaning-generating systems. Religious traditions provide narratives, categories, and evaluative frameworks that shape how people interpret the world and their place within it (Harari, 2014). AI systems increasingly generate text, images, classifications, and recommendations that influence how individuals and institutions make sense of events, identities, and decisions (Harari, 2024a).

Importantly, neither system requires comprehension or intention to shape meaning. Religious systems generate meaning because communities treat their narratives and norms as authoritative; AI systems shape meaning because users and institutions integrate their outputs into interpretive practices. This aligns with broader analyses of how information systems and technological infrastructures mediate meaning and social reality (Floridi, 2014; Verbeek, 2011). Singler's work further shows that cultural narratives surrounding AI amplify this meaning-making role by framing AI within mythic and symbolic structures (Singler, 2020).

### 4.4 Structural parallels and conceptual implications

These three features, distributed organisation, evolutionary dynamics, and meaning generation, provide the basis for a structural analogy between religion and AI. The analogy is not metaphoric but naturalistic: both systems can be understood as collective cognitive ecologies that produce and regulate shared meaning through distributed processes.

This framework has several implications:

- AI as meaning infrastructure: It reframes AI not as an autonomous agent but as a meaning infrastructure, akin to other large-scale cultural systems.
- Religion as distributed cognition**:** It situates religion within a broader class of distributed cognitive systems, highlighting its continuity with other naturalistic phenomena.
- Meaning and social life: It clarifies why AI is increasingly entangled with symbolic, ethical, and political concerns: systems that shape meaning inevitably shape social life.
- Coexisting meaning systems: It provides a conceptual basis for analysing how AI and religion may interact, converge, or conflict as coexisting meaning systems.

These implications set the stage for Section 5, where we develop the unified model that formalises these structural parallels.

### 4.5 Methodological implications

This framework also carries methodological implications for studying both religion and AI. It encourages:

- Avoiding anthropomorphism: Since neither system requires agency, intention, or comprehension to generate meaningful outputs.
- Avoiding reductionism: Religion is not reducible to cognitive biases, nor is AI reducible to engineering artefacts; both are multi-level socio-technical systems.
- System-level analysis: Focusing on distributed processes, feedback loops, and emergent patterns rather than individual agents.

- Cross-domain conceptual clarity: Providing a shared vocabulary for comparing meaning-making systems without collapsing their differences.

This methodological orientation supports a more precise analysis of how AI and religion coexist, interact, and shape contemporary cultural life.

## 5. Unified Model: Religion and AI as Distributed Meaning Systems

This section develops the paper's central theoretical contribution by articulating a unified model through which religion and AI can be understood as distributed meaning systems. The model identifies five analytically separable dimensions, evolutionary, cognitive, intersubjective, technological, and anthropological, that together illuminate the structural features shared by these otherwise distinct domains. These dimensions do not represent discrete entities or ontological layers; rather, they function as explanatory perspectives that clarify how complex systems emerge, organise themselves, and shape human interpretation and coordination. Building on the naturalistic accounts established in the preceding sections, the aim here is to synthesise those insights into a coherent conceptual architecture that supports the analysis developed in Section 6.

### 5.1 Evolutionary Layer

The first dimension concerns the historical processes through which complex systems acquire stability and functional organisation over time. As established in earlier sections, both religious traditions and contemporary AI systems develop through extended sequences of modification, retention, and refinement (Dennett, 2006; Henrich, 2015; Mallaby, 2026). What matters for the present framework is not the specific mechanisms involved, cultural transmission in one case, computational optimisation in the other, but the structural feature they share.

In both domains, organised patterns of behaviour emerge cumulatively, through processes that unfold across many iterations and often across many generations. No single agent, whether an individual believer or an AI engineer, designs the full system or determines its long-term trajectory. Instead, stability arises because certain patterns persist, are reproduced, and become entrenched within broader social or computational environments. This evolutionary dimension provides the temporal and developmental backdrop for the model as a whole, highlighting that religion and AI can both be understood as historically contingent systems shaped by iterative processes rather than by singular acts of design or intention.

### 5.2 Cognitive Dimension

The second dimension concerns the distributed cognitive organisation characteristic of both religion and AI. As established earlier, neither domain can be understood through the mental states of individual agents alone. Religious cognition is distributed across texts, rituals, institutional practices, and communal expectations, forming what Dennett describes as a system whose functional organisation exceeds the comprehension of any single participant (Dennett, 2006). This aligns with broader theories of distributed cognition, which emphasise that cognitive organisation emerges from interactions among multiple components rather than from a centralised locus of understanding (Hutchins, 1995).

Contemporary AI systems exhibit a comparable structural feature: their behaviour emerges from interactions among vast numbers of parameters, training examples, and computational operations, none of which individually encode the system's overall capacities (Agüera y Arcas, 2022; Aguera y Arcas, 2025). For the purposes of the unified model, the key point is that both religion and AI instantiate forms of distributed intelligence, where meaningful behaviour arises from the coordinated activity of many simple components rather than from a single, fully informed agent. This cognitive dimension clarifies why both systems can generate coherent outputs despite the absence of a centralised locus of comprehension.

### 5.3 Intersubjective Dimension

The third dimension concerns the shared symbolic environments that both religion and AI help to constitute. Religious traditions generate intersubjective realities, systems of meaning, value, and narrative that structure how communities interpret the world and their place within it (Harari, 2014). This aligns with social ontology, which analyses how shared symbolic structures generate institutional facts and normative expectations(Searle, 1995).

AI systems increasingly participate in similar processes by producing classifications, explanations, and representations that shape how individuals and institutions make sense of events, identities, and decisions (Harari, 2024a). The relevance of this dimension for the unified model is that both religion and AI contribute to the construction and regulation of shared meaning. They do so not by imposing fixed interpretations but by providing frameworks, categories, and outputs that people use in their interpretive practices. This intersubjective dimension highlights the extent to which both systems influence collective understanding, even in the absence of intentional meaning-making on their part.

### 5.4 Technological Dimension

The fourth dimension identifies the material and institutional scaffolding that enables both religion and AI to persist, scale, and exert influence. Religious traditions rely on cultural technologies, writing systems, ritual forms, architectural spaces, and institutional structures that stabilise meaning across time and social contexts. AI systems depend on computational infrastructures, data pipelines, optimisation regimes, hardware scaling, and organisational practices that support the emergence and maintenance of their capabilities (Mallaby, 2026).

This dimension aligns with broader analyses of technological mediation, which emphasise how technologies shape human perception, interpretation, and action (Verbeek, 2011). Within the unified model, the technological dimension emphasises that both religion and AI are embedded in and sustained by socio-material environments that extend their reach and durability. These environments do not merely support the systems; they actively shape the kinds of meaning and coordination the systems can produce.

### 5.5 Anthropological Dimension

The fifth dimension concerns the cultural narratives and symbolic interpretations that surround both religion and AI. Religious traditions are embedded in mythic and ritual structures that shape how communities understand their significance. AI systems, as Singler shows, are increasingly framed through mythic tropes, transcendence, prophecy, and existential threat, which influence how people interpret their capabilities and social roles (Singler, 2020). These narratives intersect with broader sociotechnical imaginaries that shape how societies envision technological futures and the moral significance of emerging systems (Jasanoff & Kim, 2015).

For the unified model, the anthropological dimension highlights that both religion and AI are interpreted through culturally mediated narratives that exceed their technical or doctrinal content. These narratives influence expectations, fears, aspirations, and public discourse, thereby shaping how the systems function within social life. This dimension underscores the symbolic and imaginative contexts in which both systems operate.

### 5.6 Integrated Model

Taken together, these five dimensions provide a coherent conceptual architecture for analysing religion and AI as distributed meaning systems. The evolutionary dimension situates both within long-term processes of development; the cognitive dimension explains how their functional organisation emerges from distributed structures; the intersubjective dimension clarifies their role in shaping shared meaning; the technological dimension identifies the material scaffolding that sustains them; and the anthropological dimension captures the cultural narratives through which they are interpreted.

The unified model does not claim that religion and AI are equivalent phenomena. Rather, it shows that they can be understood through a shared naturalistic vocabulary that illuminates structural parallels while preserving important differences. This framework provides the basis for the analysis developed in Section 6, where the model's explanatory power is examined in detail.

## 6. Structural Parallels and Divergences

This section examines how the unified model developed in Section 5 clarifies both the parallels and the differences between religion and AI. The aim is not to equate the two domains but to show how a shared naturalistic vocabulary illuminates their respective dynamics. The analysis proceeds dimension by dimension, drawing on the theoretical foundations established in Sections 2–4.

### 6.1 Parallels in Meaning-Making

Both religion and AI participate in the production and regulation of shared meaning. Religious traditions generate interpretive frameworks, narratives, categories, and symbolic structures that shape how communities understand the world (Harari, 2014). AI systems increasingly contribute to similar processes by producing classifications, summaries, and representations that influence how individuals and institutions interpret events (Floridi, 2014; Harari, 2024b).

The parallel lies in their systemic role as meaning infrastructures. The divergence lies in their internal logic: religious meaning is normatively thick and historically embedded, whereas AI-mediated meaning is statistically derived and normatively thin.

### 6.2 Parallels in Authority Formation

Both religion and AI generate forms of epistemic authority. Religious authority emerges from institutional structures, interpretive traditions, and claims to moral or metaphysical insight (Dennett, 2006). AI systems acquire authority through perceived objectivity, technical sophistication, and institutional reliance (Mallaby, 2026). This aligns with analyses of algorithmic authority, where trust is conferred on systems because of their computational form rather than their intentional states (Mittelstadt et al., 2016)

The parallel is that both become trusted sources of interpretation despite their internal opacity. The divergence is that religious authority is embedded in communal norms, whereas AI authority is grounded in performance metrics and the cultural prestige of computation.

### 6.3 Parallels in Coordination

Both religion and AI contribute to large-scale coordination. Religious rituals, norms, and narratives structure collective action by providing shared expectations (Henrich, 2015). AI systems coordinate behaviour through algorithmic cues, recommendations, and classifications that shape everyday decision-making. These dynamics intersect with broader sociotechnical imaginaries, which influence how societies envision and organise technological futures (Jasanoff & Kim, 2015).

The parallel is that both function as coordination mechanisms. The divergence is that religious coordination is explicit and normatively articulated, whereas AI-mediated coordination is often implicit and embedded in interfaces.

### 6.4 Parallels in Emergence and Unpredictability

Both religion and AI exhibit emergent properties that exceed the intentions of their creators. Religious traditions evolve in ways that no founding figure can fully anticipate (Henrich, 2015). AI systems develop behaviours that are not always predictable from their training data or design specifications (Mallaby, 2026).

The parallel is their emergent complexity. The divergence is temporal: religious emergence unfolds over centuries, whereas AI emergence can occur rapidly within computational environments.

### 6.5 Divergences in Ontology and Normativity

Despite structural parallels, religion and AI differ fundamentally in their ontological and normative character. Religious systems involve claims about meaning, value, and reality that are embedded in communal practices and sustained through shared symbolic frameworks (Harari, 2014; Searle, 1995). By contrast, AI systems do not possess intrinsic normativity; any normative force they appear to exert arises from the interpretive contexts in which their outputs are used. This is consistent with naturalistic accounts of meaning that emphasise the absence of inherent intentionality in computational systems (Dennett, 2006).

This divergence is central to maintaining conceptual discipline: religion is normatively saturated, whereas AI is normatively dependent.

### 6.6 Divergences in Agency and Intentionality

A further divergence concerns agency. Religious systems often attribute agency, divine, institutional, or communal, to the sources of their meaning. AI systems do not possess agency or intentionality; they operate through statistical patterning and optimisation processes (Agüera y Arcas, 2022; Aguera y Arcas, 2025). Any appearance of agency is a projection arising from human interpretive habits, a dynamic well-documented in anthropological analyses of AI myth-making (Singler, 2020).

This divergence reinforces the naturalistic framing: the model does not anthropomorphise AI nor reduce religion to computation.

### 6.7 Limits of the Analogy

Several limits are essential to maintain conceptual discipline:

- Religious meaning is historically embedded; AI-mediated meaning is computationally generated.
- Religious authority is normatively grounded; AI authority is institutionally conferred.
- Religious systems involve interpretive negotiation; AI outputs are often treated as fixed.
- Religious emergence is slow; AI emergence is rapid.

These limits ensure the model remains explanatory rather than reductive.

### 6.8 Summary

The structural parallels and divergences identified here demonstrate the explanatory power of the unified model. Religion and AI share features that allow them to be analysed as distributed meaning systems, yet they differ in ways that preserve their conceptual distinctiveness. This analysis supports the broader argument of the paper: that a naturalistic, systems-level approach provides a productive framework for understanding how emerging technologies intersect with long-standing cultural forms.

## 7. Implications

The unified model developed in Section 5, together with the analysis of structural parallels and divergences in Section 6, generates several implications for how religion and AI should be understood within contemporary philosophical inquiry. These implications concern theoretical framing, methodological orientation, and conceptual clarity. They follow directly from the naturalistic accounts of cultural evolution, distributed cognition, intersubjective meaning, and socio-technical embedding discussed earlier.

### 7.1 Theoretical Implications

#### 7.1.1 Reframing religion as a distributed meaning system

The model reinforces naturalistic approaches that treat religion as an emergent cultural formation rather than a set of propositional commitments. This aligns with accounts that emphasise the cumulative, distributed, and historically contingent nature of religious systems (Dennett, 2006; Henrich, 2015; Hutchins, 1995). The implication is that religious meaning cannot be reduced to individual belief but must be analysed at the level of system-level organisation.

### 7.1.2 Reframing AI as a socio-technical meaning system

The model clarifies that AI systems derive their significance not from computational properties alone but from their embedding within social, institutional, and interpretive environments (Harari, 2024a; Mallaby, 2026). This aligns with analyses of technological mediation, which emphasise how technologies shape human perception, interpretation, and action (Verbeek, 2011). The implication is that debates about AI should shift away from questions of artificial agency or consciousness and toward the socio-technical conditions under which AI systems shape meaning and behaviour.

### 7.1.3 Reframing the relation between the two domains

The model shows that parallels between religion and AI arise from shared structural features, distributed cognition, intersubjective influence, and emergent organisation, rather than from superficial analogy. This supports comparative work that is conceptually disciplined and avoids both technological exceptionalism and reductive equivalence.

## 7.2 Methodological Implications

### 7.2.1 A systems-level approach

The model demonstrates that both religion and AI require analysis at the level of distributed systems rather than isolated agents or artefacts. This supports methodological approaches that integrate cultural evolution, cognitive science, anthropology, and philosophy of technology (Dennett, 2006; Henrich, 2015; Verbeek, 2011).

### 7.2.2 Avoiding category mistakes

By distinguishing structural parallels from ontological divergences, the model helps prevent category mistakes, such as attributing agency to AI systems or treating religious traditions as computational analogues. This aligns with naturalistic accounts that caution against projecting intentionality onto non-agentive systems (Agüera y Arcas, 2022; Aguera y Arcas, 2025; Singler, 2020). It also resonates with analyses of algorithmic authority, which show how systems can be treated as authoritative without possessing agency (Mittelstadt et al., 2016).

### 7.2.3 Interdisciplinary coherence

The model provides a shared conceptual vocabulary that can be used across disciplines without collapsing their differences. It allows scholars of religion, AI, anthropology, and philosophy to engage with one another while maintaining conceptual clarity and avoiding cross-domain distortions. This is consistent with work on sociotechnical imaginaries, which highlights how shared conceptual frameworks shape interdisciplinary inquiry (Jasanoff & Kim, 2015).

## 7.3 Conceptual Implications

### 7.3.1 Meaning

The model suggests that meaning is not a property of individual agents alone but a product of distributed systems. This is consistent with accounts of intersubjective meaning and imagined orders (Harari, 2014; Searle, 1995) as well as naturalistic accounts of functional organisation (Dennett, 2006).

### 7.3.2 Authority

The analysis shows that authority can emerge from systems that lack intrinsic intentionality. Religious authority arises from institutional and symbolic structures; AI authority arises from institutional reliance, perceived objectivity, and the cultural prestige of computation (Mallaby, 2026; Mittelstadt et al., 2016). This challenges traditional accounts that tie authority to agency or moral standing.

### 7.3.3 Agency and intentionality

The divergences identified in Section 6 clarify that AI systems do not possess agency or intentionality, even though they can shape behaviour. This aligns with accounts that emphasise the non-agentive nature of statistical systems (Agüera y Arcas, 2022; Aguera y Arcas, 2025) and the cultural projection of agency onto technological artefacts (Singler, 2020).

### 7.3.4 Normativity

The model highlights the difference between normatively saturated systems (religion) and normatively dependent systems (AI). This distinction is grounded in the contrast between culturally embedded moral frameworks (Harari, 2014) and naturalistic accounts of meaning without intrinsic normativity (Dennett, 2006; Floridi, 2014).

### 7.4 Summary

The implications of the unified model extend beyond the comparison of religion and AI. They reshape how both domains are conceptualised, how they should be studied, and how key philosophical concepts, meaning, authority, agency, and normativity, should be understood in light of emerging socio-technical systems. These implications set the stage for the concluding section, which reflects on the broader significance of the model for contemporary philosophy of technology.

## 8. Conclusion

This paper has developed a unified model for understanding religion and AI as distributed meaning systems, grounded in naturalistic accounts of cultural evolution, cognition, and socio-technical organisation. By analysing both domains through five dimensions, evolutionary, cognitive, intersubjective, technological, and anthropological, the model provides a conceptual vocabulary that clarifies how structurally distinct systems can nonetheless exhibit comparable patterns of emergence, coordination, and influence. The analysis in Section 6 demonstrated that these parallels do not imply equivalence. Instead, they reveal how different forms of organised complexity can occupy similar functional roles in shaping human interpretation and behaviour, while remaining ontologically and normatively distinct.

The implications of this framework, outlined in Section 7, extend beyond the comparison itself. They suggest that debates about AI should not be confined to questions of artificial agency or consciousness, nor should analyses of religion be restricted to doctrinal or propositional content. Both domains require a systems-level approach that attends to distributed organisation, emergent properties, and the socio-material environments in which meaning is produced and sustained (Hutchins, 1995; Verbeek, 2011). This perspective also reframes key philosophical concepts, meaning, authority, agency, normativity in ways that better reflect the dynamics of contemporary socio-technical life (Floridi, 2014; Mittelstadt et al., 2016; Singler, 2020).

The broader significance of the model lies in its capacity to bridge domains that are often treated as incommensurable. Rather than approaching religion and AI as opposites, one ancient and symbolic, the other modern and computational, the model shows how both can be analysed through a shared naturalistic lens that highlights their structural features without collapsing their differences. This approach offers a way to understand emerging technologies not as unprecedented ruptures but as new participants in long-standing processes through which humans construct, negotiate, and inhabit shared worlds(Jasanoff & Kim, 2015).

In this sense, the model contributes to ongoing philosophical efforts to situate AI within the wider ecology of human meaning-making. It suggests that the challenges posed by AI are not solely technical or ethical but conceptual: they require rethinking how distributed systems shape interpretation, coordination, and authority in contemporary life. By providing a framework for this analysis, the paper aims to support further inquiry into the cultural, cognitive, and technological conditions under which meaning is produced today, and to clarify how emerging systems intersect with, and differ from, the cultural formations that have long structured human experience.